\documentclass[sigconf]{acmart}
\renewcommand\footnotetextcopyrightpermission[1]{} %
\AtBeginDocument{%
  \providecommand\BibTeX{{%
    \normalfont B\kern-0.5em{\scshape i\kern-0.25em b}\kern-0.8em\TeX}}}

\usepackage{framed}
\usepackage[whole]{bxcjkjatype}

\usepackage{booktabs}
\usepackage{xurl}
\usepackage{tikz}

\makeatletter
\newcommand\notsotiny{\@setfontsize\notsotiny\@vipt\@viipt}
\makeatother

\usepackage{hyperref}
\hypersetup{
    colorlinks=true,
    citecolor=red,
    linkcolor=green,
    filecolor=magenta,      
    urlcolor=blue,
    pdftitle={Canary in Twitter Mine: Collecting Phishing Reports from Experts and Non-Experts},
    pdfpagemode=FullScreen,
    }

\title{Canary in Twitter Mine: Collecting Phishing Reports from Experts and Non-experts}

\makeatletter
\def\@copyrightspace{\relax}
\makeatother
\begin{document}

\settopmatter{authorsperrow=4}

\author{Hiroki Nakano}
\email{hi.nakano.sec@gmail.com}
\affiliation{%
\fontsize{9pt}{10pt}\selectfont
\institution{NTT Security Japan (KK) \& Yokohama National University}
\fontsize{10pt}{10pt}\selectfont
\country{Japan}
}

\author{Daiki Chiba}
\email{daiki.chiba@ieee.org}
\affiliation{%
\institution{NTT Security Japan (KK)}
\country{Japan}
}

\author{Takashi Koide}
\email{takashi.koide@global.ntt}
\affiliation{%
\institution{NTT Security Japan (KK)}
\country{Japan}
}

\author{Naoki Fukushi}
\email{naoki.fukushi@global.ntt}
\affiliation{%
\institution{NTT Security Japan (KK)}
\country{Japan}
}

\author{Takeshi Yagi}
\affiliation{%
\institution{NTT}
\country{Japan}
}

\author{Takeo Hariu}
\affiliation{%
\institution{NTT Security Japan (KK)}
\country{Japan}
}

\author{Katsunari Yoshioka}
\affiliation{%
\fontsize{9pt}{10pt}\selectfont
\institution{Yokohama National University}
\fontsize{10pt}{10pt}\selectfont
\country{Japan}
}

\author{Tsutomu Matsumoto}
\affiliation{%
\fontsize{9pt}{10pt}\selectfont
\institution{Yokohama National University}
\fontsize{10pt}{10pt}\selectfont
\country{Japan}
}

\begin{abstract}
The rise in phishing attacks via e-mail and short message service (SMS) has not slowed down at all.
The first thing we need to do to combat the ever-increasing number of phishing attacks is to collect and characterize more phishing cases that reach end users. 
Without understanding these characteristics, anti-phishing countermeasures cannot evolve.
In this study, we propose an approach using Twitter as a new observation point to immediately collect and characterize phishing cases via e-mail and SMS that evade countermeasures and reach users.
Specifically, we propose CrowdCanary, a system capable of structurally and accurately extracting phishing information (e.g., URLs and domains) from tweets about phishing by users who have actually discovered or encountered it.
In our three months of live operation, CrowdCanary identified 35,432 phishing URLs out of 38,935 phishing reports, 31,960 (90.2\%) of these phishing URLs were later detected by the anti-virus engine.
We analyzed users who shared phishing threats by categorizing them into two groups: experts and non-experts.
As a results, we discovered that CrowdCanary extracts non-expert report-specific information, like company brand name in tweets, phishing attack details from tweet images, and pre-redirect landing page information.
\end{abstract}


\maketitle

\settopmatter{printfolios=true} %

\section{Introduction}
A phishing attack is an effort by an attacker to convince a user that a malicious site is legitimate to obtain information of economic value, such as account or credit card information.
Recently, phishing attacks have increased globally~\cite{liu2021detecting,reaves2016sending,safetydetectives,srinivasan2016understanding}.
In addition to the traditional phishing attacks via e-mail and short message service (SMS) have been especially on the rise~\cite{newssky}.
Attackers are exploiting SMS features for phishing: it can be sent with a phone number, with a much smaller namespace than an email address; it can be reliably pushed to cell phone subscribers when they are in range; and SMS is used for legitimate notifications and two-factor authentication, making it impossible to ignore completely.

The first step in timely combatting this ever-increasing number of phishing attacks is to collect a wider range of phishing cases that reach end users and continue understanding their characteristics.
In fact, to that end, numerous studies have been conducted to measure and analyze phishing attacks~\cite{10.1145/3133956.3134067,phishpedia,grant2019detecting,doowon2021securityanalysis}.
The facts about phishing and the weaknesses of the countermeasures revealed by these studies at that time have helped improve the coverage of spam filters in email services (e.g., Gmail and Outlook), web browser blocklists (e.g., Google Safe Browsing~\cite{gsb} and Microsoft Defender SmartScreen~\cite{smartscreen}, threat feeds (e.g., PhishTank~\cite{phishtank} and OpenPhish~\cite{openphish}), and security analysis engines (e.g., VirusTotal~\cite{virustotal} and urlscan.io~\cite{urlscan}).

However, existing countermeasures are still insufficient when phishing messages reach end users and users encounter phishing sites.
This raises the following question for us. \emph{How can we collect phishing that reaches users bypassing existing countermeasures?}

In this study, we propose an approach that uses Twitter as a new observation point to immediately collect \emph{actual phishing situations} encountered by users that have bypassed existing countermeasures and to understand the characteristics of such phishing.
Some previous studies have also used Twitter as a source to extract \emph{non-phishing} cyberattack information (e.g., vulnerability information and malware behavior information)~\cite{alves2020follow,carl2015vulnerability,hyejin2021twiti, hyejin2020cybersecurity} and limited phishing cyberattack information (e.g., search by fixed keywords or monitor only specific users)~\cite{10.1145/3548606.3559351,hyejin2021twiti,9931666}.
Specifically, these previous studies used Twitter posts of the cyberattack information by \emph{security experts}, which allowed them to identify vulnerability information and indicator of compromises (IOCs) before they were published on the National Vulnerability Database~\cite{nvd} and VirusTotal~\cite{virustotal}.
While at first glance these studies appear to be close to what our study aims to do, they differ significantly in that our goal is to extract and analyze \emph{phishing}-related information even from the actual situations that reach \emph{non-experts}.
Indeed a large number of non-experts have posted suspicious phishing attack-related cases on Twitter as alerts~\cite{welivesecurity}.
We are eager to immediately analyze the content of alerts they report as cases where phishing has reached users because existing countermeasures have been bypassed.
These reports have the benefit of being more victim-centered and comprehensive than posts by security experts and potentially being used as new information for anti-phishing technology.
Our challenge is to extract only phishing attack reports from a large number of irrelevant tweets in their everyday lives.

To this end, we propose CrowdCanary, a system capable of structurally and accurately extracting phishing information (e.g., URLs and domains) from tweets of experts and non-experts who have actually discovered or encountered phishing.
CrowdCanary is a system that employs pre-selected keywords (e.g., phishing and scam) as input to identify and output phishing attack-related user reports.
Additionally, CrowdCanary can collect a diverse set of tweets by automatically identifying and extracting new keywords that are often seen in such reports and adding them to the system.
We evaluate the effectiveness of our malicious URL collection in CrowdCanary against security engines~\cite{virustotal}, as well as existing systems that collect attack information from Twitter~\cite{10.1145/3548606.3559351,twitteriochunter}.
We also analyzed the differences between experts and non-experts and considered what approach should be taken to collect the information shared by non-experts.
Finally, we discussed how the phishing information extracted by CrowdCanary could be analyzed to help protect actual end users.

Our primary contributions are as follows.
\begin{itemize}
    \item We proposed CrowdCanary, a system that identifies reports of phishing attacks by both English and Japanese Twitter users with a high accuracy rate of 95\% for evaluation data. 
    \item We operated CrowdCanary for three months and were able to identify 38,935 phishing reports out of 19 million tweets and extract 35,432 phishing URLs. We confirmed that 31,960 (90.2\%) of these phishing URLs were later detected by anti-virus engines, demonstrating the high accuracy of CrowdCanary's threat intelligence extraction.
    \item We analyzed users who shared phishing reports and discovered that the majority of phishing reports detected by CrowdCanary were shared by non-experts. We showed that the threat intelligence reported by non-experts includes many URLs not included in the intelligence shared by experts, making it useful as a new observation point for phishing attacks from a more victim-friendly perspective.
\end{itemize}

\section{Motivating Examples}
\label{sec:motivation_challenges}
In this section, we discuss examples of user-reported phishing attacks and the challenges of extracting URLs and domain names related to phishing attacks.
\subsection{Reports on Phishing Message}

\begin{figure}[!t]
    \centering
        \includegraphics[scale=0.85]{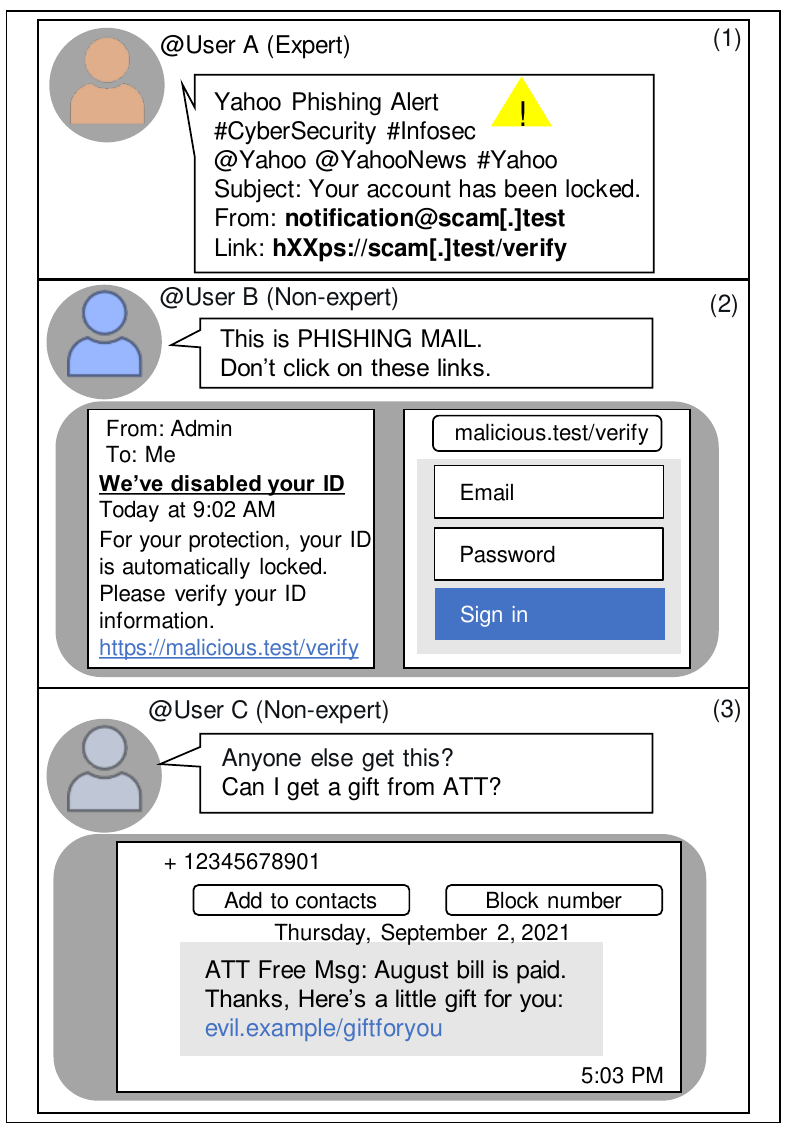}
        \caption{Reports on Phishing Messages}
        \label{fig:phishalert}
\end{figure}

With the increased usage of social media platforms and smartphones, people post phishing emails and SMSs content they discover or encounter~\cite{welivesecurity}.
Figures~\ref{fig:phishalert} (1), (2), and (3) show reports of phishing attacks posted by users on Twitter, which we refer to as cases (1), (2), and (3), respectively.
These are examples where Twitter users discover or encounter a phishing email or SMS and share that information along with the tweet's text or a screenshot taken with their smartphone.

In case (1), a user discovered Yahoo phishing emails. He/she used hashtags and mentions to alert Twitter users to the email title, the sender's email address, and the phishing URL.
It's relatively easy for us to collect reports and extract information if the report includes alerting hashtags or mentions the company's official account, and if the threat intelligence is in the body of the tweet.
In case (2), a user clicks on a URL in a phishing email, understands that he/she has arrived at a phishing site, and shares a screenshot of the email and his/her browser.
You can find the URL and domain name related to the phishing site in the information.
In case (3), a user shares a phishing SMS he/she received to get feedback because he/she are unsure if the information is real or fake.
In addition to the URL in the SMS, the text of the tweet and SMS contains the company string ``ATT,'' which was abused in the phishing attack.
Compared to case (2), this case lacks keywords such as ``PHISHING''. 
Therefore, to collect such phishing reports, we need to monitor Twitter at the right time and with the relevant keywords.
Specifically, we need a system that can extract the keyword ``ATT'' when phishing attacks with context related to ``ATT'' are prevalent and promptly collect phishing reports from Twitter using that keyword.
We will have important information about phishing attacks if we can extract URLs, domain names, and exploited company brand names as character strings from collected reports.
Because this information is based on live phishing attacks that bypassed existing countermeasure technologies and reached end users, it is valuable to consider better countermeasure technologies to detect and prevent phishing attacks before they reach users.

\subsection{Challenges}
Collecting phishing-related posts from users and extracting only phishing-related information from them presents three challenges.

\noindent\textbf{Collection of posts from various users on Twitter.}
There are a lot of tweets on Twitter, including phishing reports from security experts and non-experts.
To examine them realistically, we need to collect the tweets as narrowly as possible.
However, keywords commonly used by security experts in their reports, such as ``\#phishing'', are not always included in the reports of security non-experts.
Therefore, we need to dynamically determine keywords to include in phishing reports and collect tweets at the right time to collect reports from a wide range of users.

\noindent\textbf{Extraction of information from collected user posts.}
Phishing reports from non-experts are often presented in more diverse formats than those used by security experts.
For example, phishing-related information may only be included in the image of a tweet, not in the body of the tweet.
Without human intervention, it is difficult to determine whether the tweet is a report related to sharing information about phishing attacks from texts and images.
Since we cannot manually analyze all tweets, we need a mechanical way to extract information from both the texts and images of a large set of tweets.

\noindent\textbf{Validation of extracted information.}
It is necessary to extract only information about URLs and domain names related to phishing attacks from user reports.
Some of the information we collect may be user-generated misinformation about legitimate sites or entirely unrelated to phishing attacks.
As a result, we need to confirm the accuracy of the information extracted from the texts and images of the collected reports.

\section{Proposed System: Data Collection}
\label{sec:system}

\begin{figure*}[!t]
    \centering
        \includegraphics[scale=0.21]{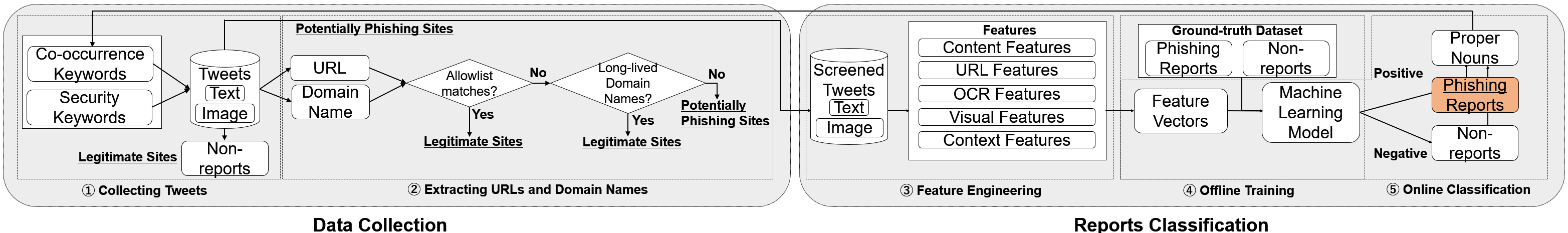}
        \caption{Overview of CrowdCanary}
        \label{fig:propoesed_system}
\end{figure*}

We propose CrowdCanary, a system that collects large-scale reports of phishing attacks in English and Japanese from Twitter users, including experts and non-experts, and allows for structured and accurate extraction of phishing information.
Figure~\ref{fig:propoesed_system} shows an overview of CrowdCanary.
CrowdCanary has two core components, \textit{Data Collection} and \textit{Reports Classification}.
In this section, we describe the first component of CrowdCanary, \textit{Data Collection}.
This component takes keywords as input for searching tweets, collects data for \textit{Report Classification}, and outputs them at one-hour intervals.
The one-hour collection interval is a customizable system parameter.
This component is designed to collect a wide range of tweets related to phishing attacks from different users.
In addition, this component extracts information about URLs and domain names that are candidates for phishing sites from the collected tweets, and excludes information that is in a notationally invalid form or related to legitimate sites.

\subsection{Collecting Tweets}
\label{subsubsec:security_keywords}
In this step, we collect tweets using two types of keywords, \textit{Security Keywords}, which are often used to share security information, and \textit{Co-occurrence Keywords}, which co-occur with \textit{Security Keywords} only at certain times.

\begin{table}[!t]
\scriptsize
\tabcolsep=1.0mm
  \centering
    \caption{Selected Security Keywords (English)}
    \begin{tabular}{l|l} \toprule
      Keywords Related & Cyber Attack, Fake Site, Fraud, Scam, Malicious Site, \\ 
      to Security Threats &  Phishing, Opendir, Spam, Social Engineering, Smishing \\
      \midrule
      Keywords with Frequent & \#CyberCrime, \#CyberSecurity,  \#CyberThreat, \#IdentityTheft, \\ 
       Shared Security Threats & \#InformationSecurity, \#InfoSec \#EmailSecurity, \#ThreatHunting, \\
       &  \#Threat, \#Security \\
      \bottomrule
    \end{tabular}
    \label{tab:list_security_keywords}
\end{table}

\noindent\textbf{Security Keywords.}
\textit{Security Keywords} in this paper refers to keywords that are regularly posted on Twitter for cybersecurity-related information.
\textit{Security Keywords} allows us to collect tweets from security experts and tweets from non-security experts sharing phishing attacks they have discovered.
Specifically, we select multiple keywords from two perspectives: related to the attack type (e.g., phishing) and information sharing (e.g., \#infosec).
We selected the 20 security keywords in Table~\ref{tab:list_security_keywords} for the following experiments.
Based on previous researches~\cite{hyejin2021twiti,hyejin2020cybersecurity} and our preliminary study, we selected keywords most likely to be shared on Twitter for information about phishing sites.
We also selected the same number of \textit{Security Keywords} in Japanese as those translated from English.

In our preliminary study, we collected and analyzed 100,000 tweets using these common keywords (e.g., ``attack'' and ``email,'') and found that more than 95\% of the tweets were unrelated to phishing attacks.
On the other hand, we also found that most tweets related to phishing attacks contained 20 selected security keywords.
Therefore, the security keywords selected in this study are reasonable for collecting and analyzing as many reports of phishing attacks as possible from many tweets on Twitter while reducing the number of false positives.

\noindent\textbf{Co-occurrence Keywords.}
\label{subsubsec:co-occurrence_keywords}
\textit{Co-occurrence Keywords} in this paper are not directly security-related keywords, but keywords (e.g., Amazon and ATT) that co-occur with \textit{Security Keywords} at certain times and are included in non-expert tweets.
Specifically, \textit{Co-occurrence Keywords} are extracted using the following procedure.
First, we consider the tweets collected during the last period when the system is running as the \textit{Co-occurrence Keywords} extraction target.
The strength of association (SoA) is then calculated using the idea of pointwise mutual information (PMI).
We define P(X) and P(Y) as the probability of the occurrence of a proper noun X and a proper noun Y, respectively, in a given tweet.
The probability that X and Y co-occur is P(X, Y).
Then, given a pair of proper nouns W in a tweet and a binary label L in the tweet (i.e., a phishing report or non-report).
In this case, PMI and SoA are represented by the following (\ref{formula:pmi}) and (\ref{formula:soa}):
\begin{align}
\label{formula:pmi}
\textstyle{PMI(X,Y) = \log (P(X,Y)/P(X)P(Y))} \\ \textstyle{SoA(W,L) = PMI(W,L) - PMI(W,\neg L)}
\label{formula:soa}
\end{align}
If X and Y do not occur at all in a single tweet, the PMI will be 0.
If X and Y are likely to occur in a single tweet, the PMI will be positive or negative.
If W appears only in phishing reports or non-reports, $PMI(W, \neg L)$ is zero, then SoA is equal to PMI ($SoA(W, L) = PMI(W, L)$).
Furthermore, W, which appears frequently in phishing and non-reports, has $PMI(W, L)$ and $PMI(W,\neg L)$ almost equal.
As a result, $SoA(W, L)$ takes on a value close to zero.
In other words, given a proper noun in a tweet for a given time period and a binary label of a phishing report or not, it is possible to extract keywords that are frequently found only in the user's report for that time period.
Since the common duration of the same phishing attack is 21 hours~\cite{adam2020sunrise}, we calculate the PMI for tweets within the previous 21 hours in our study.
For the proper noun extraction task, we use the English model~\cite{akbik2018coling} and the Japanese model~\cite{ginza}, which have been pre-trained on sufficient data and confirmed to be highly accurate for this task.
After considering whether the SoA threshold was sufficient to extract enough keywords (e.g., Amazon, ATT, Microsoft 365) related to the brand names exploited in phishing as proper nouns from user reports, we set the SoA threshold to 4.
Then, the top 10 keywords that exceed the threshold are selected as \textit{Co-occurrence Keywords}.
The default state is no \textit{Co-occurrence Keywords}, and \textit{Co-occurrence Keywords} will be selected each time this step is performed.

\subsection{Extracting URLs and Domain Names}
This step extracts URLs and domain names potentially associated with phishing attacks from the collected tweets.
The extraction targets include both the texts and images contained in the tweets.

\noindent\textbf{Image Analysis.}
We extract URLs and domain names from the images in the collected tweets by identifying the body area of the SMS or email.
Specifically, we used YOLOv5~\cite{glenn_jocher_2020_4154370} as in the previous study~\cite{10.1145/3548606.3559351}, to identify body text areas in email or SMS screenshots, annotated with 3,000 images in the dataset described in Section~\ref{subsec:dataset_creation}.
In this study, if YOLOv5 identified an area with a confidence of 0.8 or higher, we considered it to be the body text area.
Then, we use Tesseract~\cite{tesseract} to extract character strings from the body text areas in both English and Japanese.
If the body text area is not identified, we apply Tesseract to the entire image.
We extracted text from English tweets using models pre-trained in English, while we extracted text from Japanese tweets using models pre-trained in both English and Japanese.
This is because Japanese phishing emails/SMSs also contain English words.

\noindent\textbf{Text Analysis.}
Next, we extract URLs and domain names from the text of images and tweets.
Our study focuses on URLs and domain names that non-experts are likely to post as phishing attack information.
Using regular expressions, we retrieved only the matches of URLs and domain names as candidate phishing sites from both the text of tweets related to the reports and the text derived from images.
In particular, if there are defanged strings (e.g., example[.]com and hXXp) in a text, we refang the text (e.g., example[.]com to example.com and hXXp to http) and extract the URL and domain name matched by the regular expression.

\noindent\textbf{Screening Phishing-related URLs and Domain Names.}
Finally, we exclude URLs and domain names that are incorrectly formatted or related to legitimate sites.
Specifically, we check that it conforms to the format specified by RFC 3986~\cite{rfc3986} and RFC 1035~\cite{rfc1035}.
If the URL or domain name that passed format validation is not included in both the image and the text, the tweet will be excluded from further analysis.

Then, we also exclude as legitimate sites any domain name in the top 10,000 on the Tranco list~\cite{LePochat2019} and that does not match the shortened URL list~\cite{urlshorteners_github}.
Existing research~\cite{adam2020sunrise} has shown that the registration of a domain name and the execution of a phishing attack can occur within a few days or tens of days at most.
Therefore, we obtain domain name information from WHOIS and eliminate legitimate sites registered more than 365 days ago.
CrowdCanary focuses on fresher domain names to detect newer phishing attacks, thus phishing sites that are more than one year old are excluded from our study.
We output any tweets with at least one or more domain names that remain after the screening as \textit{screened tweets}.

\section{Proposed System: Reports Classification}
We describe the second component of CrowdCanary, \textit{Reports Classification}, in this section.
For the screened tweets obtained in the first component, we extract features in the tweets.
Using supervised learning, we train a classifier to identify highly relevant reports of phishing attacks with high accuracy.
From the created features, we select some features for training to achieve highly accurate and efficient classification.

\subsection{Feature Engineering}
\label{subsec:feature_engineering}
\begin{table}[!t]
\scriptsize
\tabcolsep=1.0mm
  \centering
    \caption{List of Features}
    \begin{tabular}{llllr} \toprule
    Feature Type & No. & Features Name & Vector Type & Dimensions \\ \midrule
       Content & 1 & \# of characters & Integer & 1 \\
        & 2 & \# of words & Integer & 1 \\
        & 3 & \# of hashtags & Integer & 1 \\
        & 4 & \# of images & Integer & 1 \\
        & 5 & Defanged type & Category & 9 \\ \midrule
       URL & 6 & Total \# of characters & Integer & 1 \\ 
        & 7 & \# of characters in domain name & Integer &  1 \\
        & 8 & \# of digits & Integer & 1 \\
        & 9 & Top-level domain & Category & 10 \\\midrule
       OCR & 10 & Number of characters & Integer & 1 \\ 
        & 11 & \# of words & Integer & 1 \\ 
        & 12 & \# of symbols & Integer & 1 \\ 
        & 13 & \# of digits & Integer & 1 \\ 
       \midrule
       Visual & 14 & EfficientNet Vector~\cite{pmlr-v97-tan19a} & Embedding & 16 \\ \midrule
        Context & 15 & BERT Vector~\cite{devlin-etal-2019-bert} & Embedding & 55 \\ \midrule
        Total & & & & 104 \\
     \bottomrule
    \end{tabular}
    \label{tab:features}
\end{table}

We extract features from the screened tweets that help us identify user reports.
This component classifies a single tweet as either a phishing report or a non-report.
Specifically, we generated vectorizable features from Twitter user information, tweet body text, and images.
Then, we selected helpful features from the generated features that improve the classification accuracy of phishing reports and non-reports using Boruta SHAP~\cite{boruta_shap}.
Boruta SHAP is a method that uses Shapley values for feature selection in Boruta, allowing for more accurate calculation of feature contributions and increasing the robustness of the Boruta algorithm~\cite{boruta}.
Finally, we use the five types of features shown in Table~\ref{tab:features}:  \textit{Content Features}, \textit{URL Features}, \textit{OCR Features}, \textit{Visual Features} and \textit{Context Features}.

\noindent\textbf{Content Features.}
From the content of the tweets collected in the previous component, we extract features relevant to identifying sharing related to phishing attacks, focusing mainly on the text.
Our idea is straightforward: identify the actual content of the user's tweet.
We extract five features from the information in a user's tweet.
Specifically, we designed the following five types: number of characters (No.~1), number of words (No.~2), number of hashtags (No.~3), number of images (No.~4), and defanged type (No.~5).

Features No.~1 to No.~4 are each a vector of integer values obtained from tweets.
Defanged type (No.~5) is a 9-dimensional feature vector with the one-hot encoding of 9 types of defanged types (``example .com'', ``example[.]com'', ``example(.)com'', ``example\{.\}com'', ``example\textbackslash.com'', ``hxxp://example.com'', ``hXXp://example.com'', ``ht\\tp[:]//example.com'' and ``http://example.com[/]'').
We believe that the number of characters and words in a warning-only post is relatively small.
In addition, when users post reports, they often include numerous screenshots of emails and SMSs, and these features can efficiently identify user reports.
Related studies~\cite{hyejin2021twiti,roy2021evaluating} have shown that these similar features can effectively determine whether a string contains warning information.

\noindent\textbf{URL Features.}
We extract phishing site-specific features from the URLs contained in the texts and images of the screened tweets.
Phishing sites often include characteristic strings in the domain name or path portion of the URL (e.g., abuse of subdomain names and long domain names) compared to legitimate sites~\cite{srinivasan2016understanding}.
It is possible to classify whether URLs are associated with phishing attacks by capturing the differences between the strings in the URLs of phishing sites and legitimate sites.
Specifically, we designed the following four types: total number of characters (No.~6), number of characters in the domain name (No.~7), number of digits (No.~8) and top-level domain (TLD) (No.~9).

No.~6 to No.~8 are the respective vectors of integer values calculated from the URLs (domain names) contained in the texts or images of the tweets.
We conducted a preliminary survey of the TLDs in the ground-truth dataset (Section~\ref{subsec:evaluation}) and found 841 different TLDs.
We investigated whether TLDs contribute to the identification of phishing sites using Boruta SHAP and identified 10 TLDs (``com'', ``org'', ``top'', ``info'', ``xyz'', ``online'', ``net'', ``shop'', ``cn'' and ``vip'') as important.
TLD (No.~9) is a 10-dimensional feature vector with the one-hot encoding of 10 types of TLD, as mentioned above.
For example, the fully qualified domain names (FQDNs) of phishing sites have more characters than those of legitimate sites, indicating subdomain abuse (e.g., login.security.account.example.com).
In addition, Spamhaus reports that in 2023, TLDs such as ``cn'' and ``top'' have many cases of abuse~\cite{tld_spam} and may not be reviewed by registrars.
As a result, TLDs abused by phishing sites tend to cluster in the same TLD.

\noindent\textbf{OCR Features.}
We use Tesseract~\cite{tesseract} to extract texts from the images in screened tweets.
Reports of phishing attacks shared by people in images are typically screenshots of people's smartphones, significantly different from other images commonly posted on Twitter.
We can determine if the images in the tweets are related to the report of a phishing attack by performing OCR on the images and capturing differences in the extracted strings.
If there is no image in a tweet, all OCR features are set to 0.
If a tweet has multiple images, split it, create OCR features for each image, and classify all split tweets using the same other features.

Specifically, we designed the following four types: number of characters (No.~10), number of words (No.~11), number of symbols (e.g., !, ? and \&) (No.~12) and number of digits (No.~13).
No.~10 to No.~13 are the respective integer vectors calculated from the texts extracted by applying OCR to the tweet images.
In addition to the URL and domain name, the image that the user shares as a phishing report includes the email or SMS text.
In other words, texts and words that deceive users into clicking on URLs are also included in the extracted strings.
Phishing SMSs and emails that deceive people have a predetermined amount of characters in a similar context (e.g., Your account has been suspended! Verify now [URL]), and hence the features differ significantly from strings extracted from other images.

\noindent\textbf{Visual Features.}
We construct a fixed dimensional feature vector if the tweets obtained in the previous component contain images.
Then, if there is no image in a tweet, the visual features vectors are set to 0.
If a tweet has multiple images, split it, create visual features for each image, and classify all split tweets using the same other features.
This feature captures the similarity in appearance of common phishing emails and SMSs.

Specifically, because emails, SMSs, and browser screenshots are usually images with a specific appearance, this feature is useful for classifying such images from other images.
These images are essential for distinguishing phishing reports from non-reports, as they are included when users post information in the form of images.
We use EfficientNet~\cite{pmlr-v97-tan19a} as our visual feature generation model.
We selected EfficientNet as the model for generating visual features since it is one of the state-of-the-art methods in image classification~\cite{9320487,MARQUES2020106691}.
We fine-tuned the model pre-trained on ImageNet (EfficientNet model) in English and Japanese with images related to the report (e.g., phishing email images and SMS phishing images) and images unrelated to the report (e.g., food images and landscape images).
We successfully improved the feature generation to decide whether or not to include images related to the report.

We generate a 1,280-dimensional image feature vector from tweets using a retrained model.
Then, we compressed the dimensions to achieve a cumulative contribution rate of 99\% using TruncatedSVD~\cite{hansen1987truncatedsvd}, and the result was 16 dimensions for both English and Japanese.
Here, we employ a fixed-dimensional vector, a compressed version of the vector created by the optimized EfficientNet model (No.~14).

\noindent\textbf{Context Features.}
The contextual information from the tweet sentences obtained in the previous component is represented as a fixed-dimensional feature vector.
When people share reports of phishing attacks, they often include alarming and angry statements, and are usually in a specific context.
We cannot adequately capture these contexts based on the number of characters or words in a tweet.
To this end, we use vectors created by a model trained on a large amount of text to capture the context of a tweet's text.

Specifically, we use BERT~\cite{devlin-etal-2019-bert} as the context feature generation model.
BERT and BERT-based methods are state-of-the-art for several natural language processing tasks~\cite{1904.08398,2007.01852,1901.04085}.
We fine-tuned the sentences of tweets related to reports in both English and Japanese using the ground-truth dataset (Section~\ref{subsec:dataset_creation}).
We optimized feature generation for a pre-trained model with many words to determine whether a tweet is related to user reports or not.
In certain scenarios, a user who receives a phishing attack alerts, suspects, or incites the attacker.
As a result, the contextual characteristics are different from other people's daily posts.

We create a 768-dimensional context feature vector from tweets using a retrained model.
Then we also compressed the dimensions to achieve a cumulative contribution rate of 99\% using TruncatedSVD~\cite{hansen1987truncatedsvd}, and the result was 58 dimensions for both English and Japanese.
Here, we use a fixed-dimensional vector, a compressed version of the vector generated by the optimized BERT model (No.~15).

\subsection{Training and Classification}
Using the many features we have created so far, we train a model for binary classification of whether a tweet is a report of a phishing attack or not.

\noindent\textbf{Method.}
Given labeled positive or negative training data, a supervised learning model can be trained that uses the characteristics of each tweet to predict the binary value of tweets associated with phishing reports or non-reports.
We then aim to predict with a high degree of accuracy whether new tweets are similar to previous phishing reports or non-reports.
We compared and evaluated eight commonly used supervised learning algorithms: Random Forest, Neural Network, Decision Tree, Support Vector Machine, Logistic Regression, Na\"ive Bayes, Gradient Boosting, and Stochastic Gradient Descent.
To account for the influence of some algorithms on accuracy loss, all feature vectors were preprocessed to set the mean to 0 and the variance to 1.
Here, we train and evaluate using a ground-truth dataset labeled with phishing or non-phishing reports, which will be explained later in Section~\ref{subsec:dataset_creation}.

\noindent\textbf{Results.}
We adopted Random Forest as the training and classification algorithm for the following three reasons.
(1) Random Forest showed the best binary classification accuracy for the ground-truth data among the eight algorithms.
(2) Random Forest performed consistently well with stable speed in both the training and inference phases for large amounts of data.
(3) The importance of the features in the Random Forest was distributed among \textit{Content Features}, \textit{URL Features}, \textit{OCR Features}, \textit{Visual Features}, and \textit{Context Features}, thus the classifier does not depend on any particular feature in its decision.
We perform a classification accuracy evaluation on the ground-truth datasets (Section~\ref{subsec:dataset_creation}) and, in the live operation using CrowdCanary (Section~\ref{sec:analysis}), a model trained with the Random Forest algorithm, to perform the binary classification of phishing reports and non-reports.

\subsection{Evaluation of Classification Accuracy}
\label{subsec:evaluation}
\begin{table}[!t]
\scriptsize
\tabcolsep=1.0mm
  \centering
    \caption{Ground-truth Dataset for Evaluating the Accuracy of Machine Learning Models}
    \begin{tabular}{lllr} \toprule
      Language & Collected Time & Label & \# of Tweets \\ \midrule
      English & May. 1, 2021 -- Jul. 19, 2021 & Phishing Reports & 5,000 \\ 
       & (80 days) & Non-Reports & 15,000 \\ \midrule
      Japanese & May. 1, 2021 -- Jul. 19, 2021 & Phishing Reports & 5,000 \\
       & (80 days) & Non-Reports & 15,000 \\ 
       \bottomrule
    \end{tabular}
  \label{tab:dataset_train}
\end{table}

\begin{table}[!t]
\scriptsize
\tabcolsep=0.5mm
  \centering
    \caption{Classification Accuracy Evaluation Results}
    \begin{tabular}{llrrrrr} \toprule
      Language & Features & Accuracy & TPR & TNR & Precision & F-measure\\ \midrule
      English & \textbf{Content+URL+OCR+Visual+Context} & \textbf{0.957} & \textbf{0.952} & \textbf{0.962} & \textbf{0.962} & \textbf{0.957} \\
       & Content+URL+OCR & 0.838 & 0.829 & 0.847 & 0.845 & 0.837 \\ \midrule
      Japanese & \textbf{Content+URL+OCR+Visual+Context} & \textbf{0.949} & \textbf{0.948} & \textbf{0.960} & \textbf{0.951} & \textbf{0.943} \\
       & Content+URL+OCR & 0.798 & 0.754 & 0.843 & 0.827 & 0.789 \\
      \bottomrule
    \end{tabular}
    \label{tab:results_evaluation}
\end{table}

Before taking measurements with CrowdCanary in live operation, we evaluated the classification accuracy of phishing reports and non-reports in CrowdCanary.

\noindent\textbf{Ground-truth Datasets.}
\label{subsec:dataset_creation}
Table~\ref{tab:dataset_train} shows the dataset used for the evaluation.
First, we used the 20 English keywords from Table~\ref{tab:list_security_keywords} and the 20 translated Japanese keywords.
Then, we searched on Twitter using the keywords for 80 days from May. 1, 2021 -- Jul. 19, 2021, and collected 1,543,245 and 1,023,368 tweets in English and Japanese, respectively.
Existing studies or publicly available datasets do not provide ground-truth datasets for the correct answers to phishing reports and non-reports, which are our research goals.
As a result, we have to annotate them ourselves.
Therefore, we randomly sampled the collected tweets and manually labeled them with a binary value of either phishing reports or non-reports.
We excluded from our annotations tweets that do not have a URL or domain name in the text or image of the tweet.
We then accessed the URLs and domain names in the text and images of the collected tweets from the experimental environment, examined the collected web content, and performed a similarity analysis with legitimate sites.
Four security engineers conducted this annotation, and we labeled each of the tweets that we all agreed were reports of phishing attacks and non-reports.
As a result of the annotations, we labeled the tweets as ``phishing reports'' when we determined they were related to phishing attacks and ``non-reports'' when they were not.
Finally, we created 5,000 ``phishing reports'' and 15,000 ``non-reports'' in English and Japanese, respectively.
To account for the effect of temporal bias, we split the training and testing data 7:3 in time order for the evaluation experiment.

\noindent\textbf{Evaluation Results.}
The evaluation results are shown in Table~\ref{tab:results_evaluation}.
When combining all features (Content+URL+OCR+Visual+Context) for the English case, Accuracy was 0.957, True Positive Rate (TPR) was 0.952, True Negative Rate (TNR) was 0.962, Precision was 0.962, and F-measure was 0.957.
The results show that the accuracy is sufficient to classify phishing reports from the large volume of tweets collected.
We also found that it is difficult to detect user reports of phishing attacks with high accuracy using only simple features generated from meta information on Twitter.
The same result is obtained for the Japanese case.
We conclude that feature vectors with information embedded in a fixed dimension, pre-trained on many languages and images, significantly improve classification accuracy.
To summarize, in subsequent evaluations for Section~\ref{sec:analysis}, we will use a machine learning model trained by combining five types of features: Content+URL+OCR+Visual+Context.

\section{Evaluating User Reports in the Wild}
\label{sec:analysis}
\begin{table}[!t]
\scriptsize
\tabcolsep=1.0mm
  \centering
    \caption{Overview of Datasets for Evaluation}
        \begin{tabular}{lllr} \toprule
        System & Period & Datasets & \# \\ \midrule
        CrowdCanary & Nov. 1, 2022 -- Jan. 31, 2023 & Collected Tweets & 18,765,699 \\ 
         & (3 months) & Screened Tweets & 324,589 \\ 
         & & Phishing Reports & 38,935 \\
         & & Detected Threats & 42,987 \\
         & & Detected URLs & 35,432 \\ 
        \midrule
        SpamHunter & Jan. 1, 2018 -- Aug. 31, 2022 & Detected Threats & 15,553 \\
         & (56 months) & Detected URLs & 15,269 \\
        \midrule
        Twitter IOC Hunter & Aug. 1, 2021 -- Jul. 31, 2022 & Detected Threats & 10,092 \\
         & (12 months) & Detected URLs & 9,344 \\
        \bottomrule
    \end{tabular}
    \label{data_overview}
\end{table}
We performed a comparative evaluation with two existing systems~\cite{twitteriochunter,10.1145/3548606.3559351} that collect and publish malicious URLs and domain names from Twitter.

\subsection{Datasets for Evaluation}
A summary of the datasets for CrowdCanary and the two existing systems for comparison is shown in Table~\ref{data_overview}.
These two existing systems collect information from Twitter, but the information they collect is not limited to phishing attacks.
Although CrowdCanary focuses specifically on phishing attacks, we demonstrate that the quantity and quality of information collected by CrowdCanary outperforms the two existing systems.
While CrowdCanary is a newly implemented system that works perfectly on the current version of Twitter, the existing systems rely heavily on older Twitter APIs and are unable to analyze the latest tweets.
Therefore, we used datasets~\cite{spamhunter,twitteriochunter} from when these systems were publicly available for our evaluation.

\noindent\textbf{Proposed System (CrowdCanary).}
\label{subsec:collection_results}
We ran CrowdCanary continuously every hour for three months, from Nov. 1, 2022 -- Jan. 31, 2023.
We set the \textit{Security Keywords} to 20 English and 20 Japanese words in Table~\ref{tab:list_security_keywords}, and the initial state of the \textit{Co-occurrence Keywords} to none.
CrowdCanary selected new \textit{Co-occurrence Keywords} every hour from the collected user reports.
During the three-month experiments, we collected 18,765,699 tweets, screened 324,589 tweets, and identified 38,935 phishing reports.
For domain names included in user reports, we considered them to be URLs by appending the protocol ``https'' to the domain name.
Finally, we merged these URLs with the extracted URLs to obtain 35,432 unique URLs extracted by CrowdCanary.

\noindent\textbf{Existing System (SpamHunter).}
We selected the dataset of the previous study~\cite{10.1145/3548606.3559351} as our existing system for comparison.
Their ``SpamHunter'' system collects tweets with SMS-related keywords, performs image analysis, and extracts phishing-related URLs.
Spam\\Hunter comes closest to our motivation in terms of the information we want to collect, however their method of collecting tweets is very limited.
They published the collected URLs~\cite{spamhunter}, and obtained 15,553 threats from Jan. 1, 2018 -- Aug. 31, 2022.
In addition, we added ``https'' to threats that lacked protocol information, excluded URLs with formatting deficiencies, and finally prepared 15,269 detected URLs.

\noindent\textbf{Existing System (Twitter IOC Hunter).}
Next, we selected the existing system~\cite{twitteriochunter} for comparison because it extracts cybersecurity-related information (e.g., malicious URLs, IP addresses, etc.) from Twitter and allows us to obtain data for a specified time period through its API.
We obtained 10,092 threats using the API of Twitter IOC Hunter~\cite{twitteriochunter} from Aug. 1, 2021 -- Jul. 31, 2022.
Similar to SpamHunter, we added ``https'' to threats that lacked protocol information, excluded URLs with formatting deficiencies, and finally prepared 9,344 detected URLs.
\subsection{Comparison of Maliciousness using VirusTotal}
\label{subsec:vt}
\begin{table}[!t]
\scriptsize
 \tabcolsep=1.0mm
  \centering
    \caption{Overview of Comparison Results between CrowdCanary and Existing Systems}
        \begin{tabular}{l|rrr|rr} \toprule
        System & VT$\geqq$1 & VT$\geqq$5 & Total & VT$\geqq$1 /day &  VT$\geqq$5 /day \\ \midrule
        \textbf{CrowdCanary} & \textbf{31,960} & \textbf{15,768} & \textbf{35,432} & \textbf{347} & \textbf{171} \\ %
        \textbf{(Image+Text)} & \textbf{(90.2\%)} & \textbf{(44.5\%)} & \textbf{(100.0\%)} & \\
        CrowdCanary & 17,633 & 7,267 & 19,205 & 87.4 & 29.9  \\  %
        (Only Image) & (84.2\%) & (37.8\%) & (100.0\%) & & \\
        CrowdCanary & 15,260 & 8,452 & 17,231 & 164 & 124 \\ %
        (Only Text) & (88.6\%) & (49.1\%) &(100.0\%) & & \\
        SpamHunter & 8,266 & 1,718 & 15,269 & 4.85 & 1.01 \\ %
        \cite{10.1145/3548606.3559351} & (59.8\%) & (10.9\%) & (100.0\%) & & \\
        Twitter IOC Hunter & 5,228 & 2,172 & 9,344 & 14.3 & 5.95 \\ %
        \cite{twitteriochunter} & (56.0\%) & (23.2\%) & (100.0\%) & & \\
        \bottomrule
    \end{tabular}
    \label{tab:comparison_results_urls}
\end{table}

We analyzed how VirusTotal (VT)~\cite{virustotal} flags the URLs detected by CrowdCanary and the two existing systems~\cite{10.1145/3548606.3559351,twitteriochunter}.
When we request VirusTotal to scan a URL, it evaluates the maliciousness of about 90 different types of anti-virus software and returns the results to us.
Several studies~\cite{10.1145/3320269.3384714,10.1145/3355369.3355585,hyejin2021twiti,10.1145/3278532.3278569,zhu2018chainsmith} used VirusTotal as a metric for evaluation.
Then it is appropriate for our study to evaluate how much of the information collected from Twitter are actually malicious URLs.

VirusTotal provides five types of results for scanned URLs: malicious, suspicious, harmless, undetected and timeout.
Because CrowdCanary immediately collects/outputs phishing attacks shared by Twitter users, sometimes VirusTotal does not detect them even though the URLs are malicious.
We then requested scans and obtained results at least one week after detection in CrowdCanary.
Since the URLs of the existing systems had already mainly been analyzed by VirusTotal, we obtained the results of the scans.
If VirusTotal had no previous scan results, we requested a scan and obtained the scan results.
VirusTotal has also seen cases of false positives from anti-virus vendors~\cite{10.1145/3355369.3355585}; therefore, URLs identified as malicious/suspicious by one anti-virus vendor are not necessarily phishing URLs.
As a result, in our study, we compared URLs flagged as malicious/suspicious by at least one and five anti-virus vendors in VirusTotal with CrowdCanary and two existing systems.

The comparison results are shown in Table~\ref{tab:comparison_results_urls}.
Focusing on URLs that were flagged as positive by five or more antiviruses in VirusTotal, 15,768 (44.5\%) were positive for CrowdCanary (Image+Text), 7,267 (37.8\%) were positive for CrowdCanary (Only Image), 8,452 (49.1\%) were positive for CrowdCanary (Only Text), 1,718 (10.9\%) were positive for SpamHunter and 2,172 (23.2\%) were positive for Twitter IOC Hunter.
We confirmed that CrowdCanary was superior to the proposed and existing systems in terms of both the absolute number and detection rate of URLs later detected by multiple antiviruses in VirusTotal.
SpamHunter is a system that extracts information from tweet images, and Twitter IOC Hunter is a system that extracts threats from tweet texts.
Even if we target only images and texts for CrowdCanary's threat extraction, we can see that it can extract more URLs detected by VirusTotal.
Due to the different experimental periods of the proposed system and the two existing systems, we compared the average per day of URLs detected by VirusTotal.
In this case as well, the results showed that CrowdCanary was superior to the existing systems.
In particular, the number of URLs detected by VirusTotal in five or more anti-viruses was 171 per day for CrowdCanary, which extracts information from images and text, about 170 times higher than SpamHunter~\cite{10.1145/3548606.3559351} and about 29 times higher than Twitter IOC Hunter~\cite{twitteriochunter}.

Additionally, we manually investigated the remaining 3,472 (=35,432-31,960) URLs that VirusTotal did not detect during the experimental period.
We identified malicious URLs that could be identified as phishing sites based on the content of tweets, website content, screenshots, WHOIS information, etc.
As in Section~\ref{subsec:dataset_creation}, this investigation was conducted by four security engineers and took a total of 30 hours to check for undetected URLs in VirusTotal.
As a result, we found that 2,635 (7.44\%) URLs were truly phishing sites (false negatives by VirusTotal).
Most of these URLs were used for redirects under the domain names of duckdns.org, which abused the dynamic DNS provider, and cutt.ly, which abused the URL shortening service and made it difficult to determine the maliciousness of the URLs mechanically.
On the other hand, 482 (1.36\%) URLs were incorrect information due to OCR misidentification (e.g., misidentifying ``l'' as ``1''), 160 (0.45\%) URLs were not phishing site URLs included in the user's report (e.g., minor legitimate sites that users cannot accurately determine whether they are phishing or not), and 195 (0.56\%) URLs were misclassified by the machine learning model (e.g., legitimate SMSs or emails).
The next Section~\ref{sec:experts_non-experts} analyzes a total of 34,595 (=31,960+2,635) URLs detected by VT or manually identified as phishing URLs.

\section{Comparison of Experts and Non-experts}
\label{sec:experts_non-experts}
\begin{figure}[!t]
    \centering
        \includegraphics[scale=0.12]{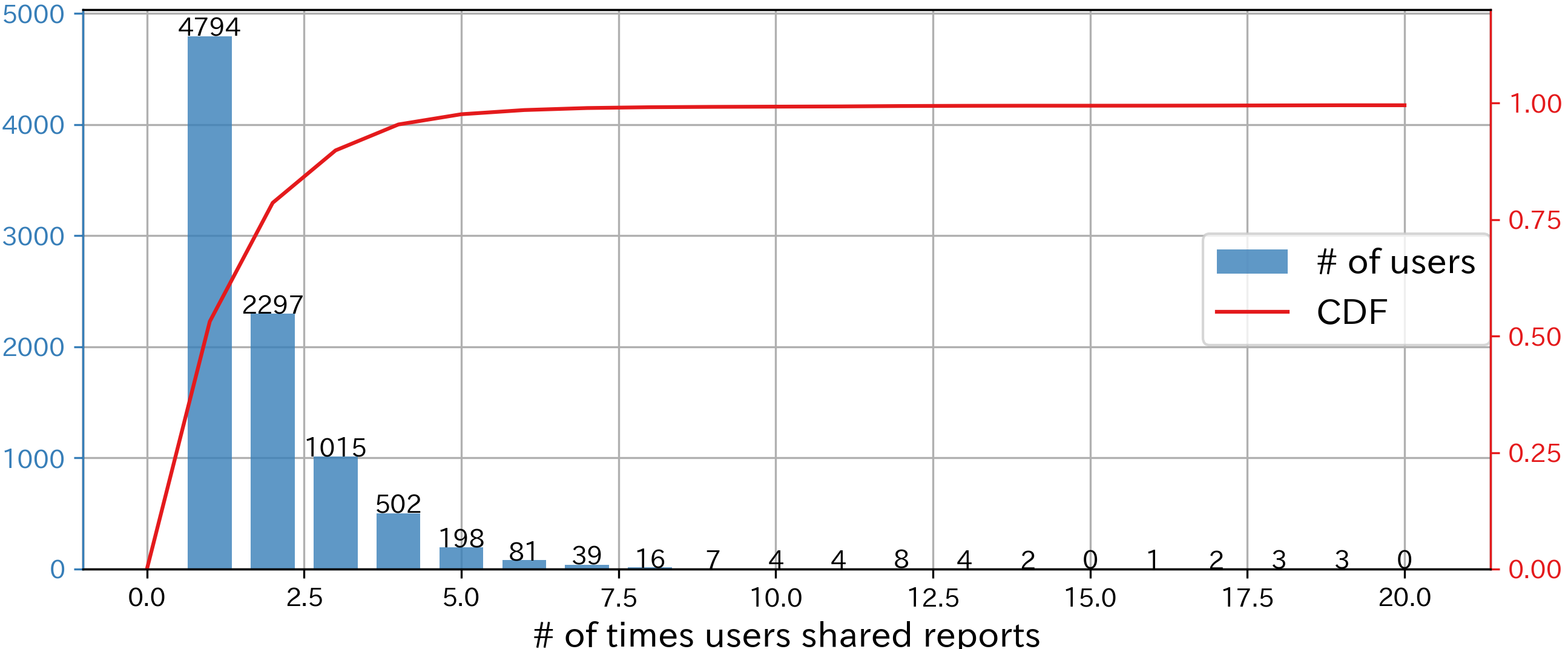}
        \caption{Correlation between Users and Number of Times Reports Were Shared}
        \label{fig:crowdcanary_reports_share}
\end{figure}

\begin{table}[!t]
\scriptsize
 \tabcolsep=1.0mm
  \centering
    \caption{User Categorization Results}
        \begin{tabular}{l|rr|rrrr} \toprule
         & & & \# Shared & & & \\
        User Type & \# Users & \# Reports & min & median & mean & max \\ \midrule
        Expert & 25 & 15,263 & 1 & 280 & 610 & 3,900 \\ 
        Non-expert & 9,000 & 17,577 & 1 & 1 & 1.95 & 73 \\ 
        \bottomrule
    \end{tabular}
    \label{tab:user_type}
\end{table}
In this section, we analyze characteristics of users (security experts or non-experts) by using 34,595 URLs (32,813 phishing reports) that contain malicious information about phishing attacks identified by VirusTotal and manual investigation in Section~\ref{subsec:vt}.

\subsection{Analysis of Users who Shared Reports}
Of the 32,813 phishing reports, the number of unique users was 9,025.
We identify the users who shared these reports as experts or non-experts.
Specifically, users who satisfy either of the following two conditions are considered experts, and users who satisfy neither of the two conditions are considered non-experts.
(1) The user has security-related keywords (e.g., phishing, threat hunter) in their Twitter profile.
(2) The user has posted more than half of their last 10 tweets related to cybersecurity.

As a result, we categorized users in the method described above, resulting in 25 users (2.77\%) as experts and 9,000 users (97.23\%) as non-experts, as shown in Table~\ref{tab:user_type}. 
We reviewed the results as a manual and verified that they were categorized as intended.
We found that experts share phishing reports an average of 610 times, while non-experts share phishing reports an average of 1.95 times.
In particular, we confirmed that many expert shares appeared to be mechanical, with some accounts only posting phishing attack threats up to 3,900 times during the experimental period.
Most non-experts shared phishing emails and SMS messages they received only a few times.
However, in rare cases, we found some non-experts who shared phishing emails and SMS messages they received 73 times during the experimental period.

Additionally, Figure~\ref{fig:crowdcanary_reports_share} shows the correlation between users and the number of times reports are shared.
The x-axis represents the number of times a user shared a report, the blue bar on the y-axis represents the number of reports based on the number of times the report was shared, and the red line on the y-axis represents the cumulative distribution function (CDF) value of the reports.
From Figure~\ref{fig:crowdcanary_reports_share}, users who shared only one report accounted for 53.1\% of the total, while users who shared two reports accounted for 78.6\% of the total.
In other words, if we collect information from Twitter limited to accounts of users who frequently share, as in existing studies~\cite{carl2015vulnerability,hyejin2021twiti}, we would miss phishing reports from numerous users.
We demonstrated that CrowdCanary can collect not only the limited information shared by security experts, but also information posted by a large number of users, including reports of phishing attacks by non-experts.
\subsection{Analysis of the detected URLs' characteristics}

\begin{figure}[!t]
    \centering
        \includegraphics[scale=0.12]{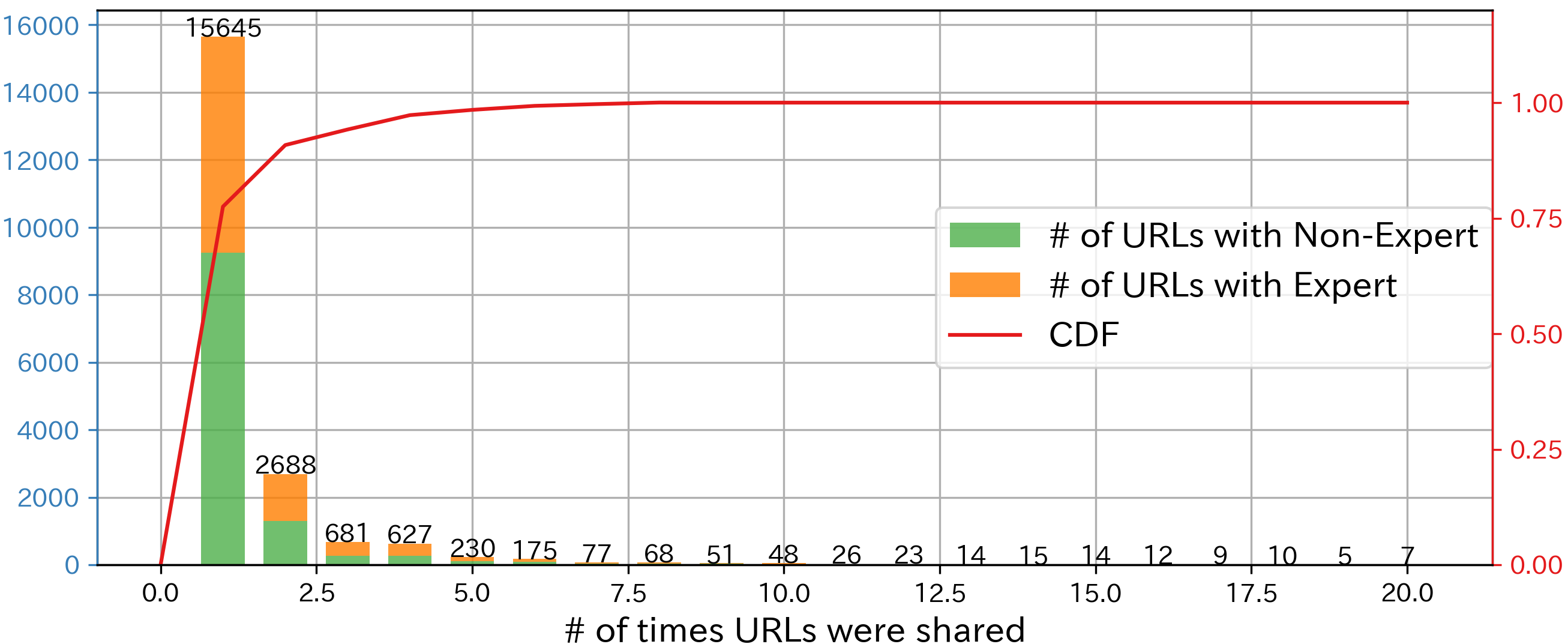}
        \caption{Correlation between Users Types and Number of Times URLs Were Shared}
        \label{fig:crowdcanary_urls_share}
\end{figure}

\begin{table}[!t]
\scriptsize
 \tabcolsep=0.9mm
  \centering
    \caption{Comparison of URLs and FQDNs Characteristics}
        \begin{tabular}{l|rrr|rrr} \toprule
        User Type & \# URLs & \# Shortened URLs & \# Services & \# FQDNs & \# Dynamic DNSs & \# Providers \\ \midrule
        Expert & 16,778 & 102 (0.61\%) & 7 & 6,530 & 668 (10.2\%) & 1 \\ 
        Non-expert & 18,654 & 2,896 (15.5\%) & 13 & 8,699 & 3,612 (41.5\%) & 3 \\ 
        \bottomrule
    \end{tabular}
    \label{tab:url_type}
\end{table}

We analyzed the value of the URLs included in the phishing reports.
Specifically, we analyzed the number of times each URL was shared as a phishing report.
The correlation between user types (i.e., experts or non-experts) and the number of times reports containing that URL were shared is shown in Figure~\ref{fig:crowdcanary_urls_share}.
The x-axis represents the number of times a URL has been shared, the green and orange bars on the y-axis represent the number of URLs found that match, and the red line represents the CDF value of the unique URLs.
From Figure~\ref{fig:crowdcanary_urls_share}, URLs extracted from phishing reports shared only once by users accounted for 77.5\% of the total, while URLs extracted from user reports shared twice by users accounted for 90.8\%.
As shown in our results, we found that extracting information from the tweets of a fixed set of users with a limited observation target would miss the majority of high-value malicious URLs that are shared a few times at most.

We then analyze the characteristics of the URLs and FQDNs shared by security experts and non-experts.
The unique URLs included in the expert and non-expert reports were 16,778 and 18,654, respectively, as shown in Table~\ref{tab:url_type}.
Attackers sometimes use redirects from the landing URL to the phishing site where they ultimately want to direct the user~\cite{adam2020sunrise,cisco}.
Specifically, we investigated how many URLs exploited the dynamic DNS providers~\cite{dynamicdns} and URL shortening services~\cite{urlshorteners_github} used to redirect phishing attacks.
Among dynamic DNS providers, duckdns.org was found to be abused 99.3\% in total, and among URL shortening services, cutt.ly and bit.ly were abused 70.5\% in total.
Because these services and providers are free, can generate a large number of URLs, and have no countermeasures to exploit for phishing attacks, it is believed that attackers use them to evade detection (i.e., spam emails and SMSs detection) of phishing sites they have created.
Many of the threats shared by non-experts are URLs that are actually spread in phishing e-mails and SMSs.
These URLs can be used as a starting point for analyzing the full picture of attacks, or as intelligence for block lists that automatically detect spam e-mails and SMSs.
Experts often share redirected URLs that, without proper referrer~\cite{zhang2021crawlphish}, are unanalyzable and inadequate for phishing email and SMS prevention.

\subsection{Analysis of Report Sharing Methods}

\begin{table}[!t]
\scriptsize
 \tabcolsep=1.0mm
  \centering
    \caption{Comparison of Report Sharing Methods}
        \begin{tabular}{l|rr|rr|rr} \toprule
         & & & \# Hashtags &  & \# Mentions & \\
        User Type & \# URLs in Images & \# URLs in Texts & median & mean & median & mean \\ \midrule
        Expert & 1,523 (10.0\%) & 13,740 (90.0\%) & 4 & 3.83 & 0 & 0.21 \\ 
        Non-expert & 16,659 (94.8\%) & 918 (5.22\%) & 0 & 0.73 & 0 & 0.15 \\ 
        \bottomrule
    \end{tabular}
    \label{tab:share_type}
\end{table}

\begin{table}[!t]
\scriptsize
 \tabcolsep=0.4mm
  \centering
    \caption{Top 10 Keywords Collected Phishing Reports}
        \begin{tabular}{l|llr|llr} \toprule
        Rank & Keywords (Expert) & Type & \# & Keywords (Non-expert) & Type & \# \\ \midrule
        1 & \#phishing (E+J) & Security & 3,982 & \#phishing (E+J) & Security & 1,545 \\ 
        2 & \#scam (E+J) & Security & 2,733 & National Tax Agency (J) & Co-occurrence & 1,175 \\ 
        3 & \#phishingmail (J) & Co-occurence & 1,406 & Fraud (J) & Security & 1,035 \\ 
        4 & \#infosec (E) & Security & 1,283 & \#Amazon (E+J) & Co-occurence & 889 \\ 
        5 & \#cybersecurity (E)  & Security & 1,280 & \#scam (E+J) & Co-occurrence & 820 \\ 
        6 & \#Amazon (E+J) & Co-occurence & 1,119 & Softbank (J) & Co-occurrence & 712 \\ 
        7 & \#phishingsite (J) & Co-occurence & 1,079 & Docomo (J) & Co-occurrence & 688 \\ 
        8 & National Tax Agency (J) & Co-occurence & 1,022 & American Express (E+J) & Co-occurrence & 653 \\ 
        9 & SMBC (J) & Co-occurence & 894 & Google (E+J) & Co-occurrence & 512 \\ 
        10 & \#bank (E) & Co-occurence & 822 &Please retweet (J) & Co-occurrence & 488 \\ 
        \bottomrule
        
    \end{tabular}
    \label{tab:keywords}
    \\ E: English only, J: Japanese only, E+J: Contains identical semantic words in both languages
\end{table}

We analyze the differences in the way experts and non-experts share information.
First, we compared experts and non-experts on how users share information about phishing attacks.
The results are shown in Table~\ref{tab:share_type}.
We found a significant difference in how information was shared: 90\% of expert reports included URL information in the text of their tweets.
In contrast, 95\% of non-expert reports included URL information in the images of their tweets.
Experts identify threats through their own investigation rather than by encountering them, and they often share the information in a formatted text (in the text of a tweet).
On the other hand, non-experts often store the phishing attacks they encounter it (receiving an email or SMS, or reaching the site with a browser) as screenshots from their smartphones, etc., and attach the images directly to their tweets and share them.
Although it is difficult to collect a large number of these reports from non-experts and extract information properly, CrowdCanary was able to extract as many threats as experts and more, as shown in Figure~\ref{fig:crowdcanary_urls_share}.

We also found significant differences in features between experts and non-experts in the context of the text when sharing reports.
The median and the mean number of hashtags and mentions in the phishing reports of experts and non-experts are shown in Table~\ref{tab:share_type}.
Hashtags are referred to as ``\#phishing'' and are primarily used by users on Twitter to share information.
People looking for information can find tweets containing the hashtag relatively easily using the search function.
In this case, the expert report shows an average of 3.83 hashtags in the tweets, while the non-expert report shows an average of only 0.73 hashtags.
As a result, collecting non-expert reports with appropriate keywords is more difficult than collecting expert reports shared using fixed hashtags.
Similarly, we examined user reports that included mentions that could be posted to a specific user account on Twitter and found no significant differences between experts and non-experts.

Finally, we discuss query keywords that were useful in collecting phishing reports.
The top 10 keywords that resulted in the collection of expert and non-expert reports are listed in Table~\ref{tab:keywords}.
Among the top 10 keywords for experts, 8 were hashtagged and 4 were security (as defined in Section~\ref{subsubsec:security_keywords}) keyword types.
In particular, we found that a large number of experts shared their information using the hashtags ``\#infosec'' and ``\#cybersecurity'', which are not commonly used by non-experts.
On the other hand, only 3 of the top 10 non-expert keywords were hashtagged.
Although ``\#phishing'' was sometimes the most effective keyword for collecting phishing reports, as it was for experts, many of the non-experts shared reports using the name of the company brand that was exploited in the phishing attack.
However, simply searching for a company's brand name will return a number of irrelevant tweets.
Therefore, either a search using appropriate keywords at the right time, as in this study, or a highly accurate detection mechanism from among the tweets continuously collected by company brand name is required.

\section{Discussion}
\label{sec:discussion}
We describes the potential for using CrowdCanary output information to defend against phishing attacks, the limitations of CrowdCanary, and ethical considerations of the experimental design.

\subsection{Utilizing the Intelligence Collected for Phishing Attack Defense}
We have demonstrated that CrowdCanary can collect threat intelligence on a large number of phishing attacks with greater accuracy than existing technologies.
How can this collected intelligence be applied to actual defensive strategies?
We believe that intelligence can be used from two main perspectives.

First, the phishing information collected can add to the intelligence in the block lists.
It has been reported that the spread of phishing attacks does not end with the first wave of attacks; the second and third waves of attacks are sometimes spread using the same domain names~\cite{chao2012analyzing}.
By extracting information about the attack as early as possible, such as during the first wave, and feeding it into blocklists (e.g., email spam filters), it may be possible to protect users who may become victims of the second and third waves.
It was also reported that among users who receive phishing emails, the average time difference between the timing of the first user to click on the URL and the last user to click on the URL is 21 hours~\cite{adam2020sunrise}.
By sharing information with the browser vendor's block list during this time difference, the browser can warn the user and protect them from phishing attacks if they visit the same URL.

Second, the characteristics of phishing attacks contained in the collected information can be analyzed and used as countermeasure information for similar attacks that may occur in the future.
It has been reported that phishing sites change domain names frequently, but may continue to be hosted at a particular IP address~\cite{10.1145/3038912.3052654}.
For example, using passive DNS (e.g., Farsight DNSDB~\cite{pdns}), it is possible to detect attacks early using CrowdCanary intelligence if the A record of a newly appearing domain name is linked to the same IP address as a domain name that has been exploited for phishing attacks in the past.
In addition, phishing sites created using phishing toolkits often have the identical HTML source, images on the site, and scripts~\cite{hugo2021phishers}.
This information can be useful in techniques such as content-based phishing site detection~\cite{xiang2011cantina+}.
In addition, information about phishing attacks received by many users can be used to understand trends in company brands being exploited in attacks and to keep an eye on companies and industries that attackers will be targeting in the future.

\subsection{Limitation}
Our study has three limitations.

First, our study does not focus on extracting reports only from information about the final destination of phishing sites that involve user interaction or redirection.
For example, some users may only share a screenshot on Twitter with the URL of the final destination after the entry or redirect occurs.
CrowdCanary cannot properly extract reports in this case because it has no information about the user's input or redirection behavior on the browser.
In particular, CrowdCanary is a system that collects URLs that are the seeds of phishing attacks.
CrowdCanary does not focus on phishing attacks that do not redirect without an acceptable referrer or can only be reached by clicking.
These attacks can be handled by crawling URLs extracted by CrowdCanary as seeds in existing researches~\cite{10.1145/3320269.3384714,191004}.

Second, the features designed in this paper are chosen to be invariant with respect to user reporting of phishing attacks.
However, the system's accuracy will inevitably decrease over time, and the system will need to be relearned each time, but this is an issue for future work.

Finally, depending on recent Twitter specification changes~\cite{twitter_dev}, equivalent information may no longer be available via the API.
Any social networking service that allows users to post photos and text, as popular as Twitter, can be used as a source of threat intelligence in the same way.
In addition, CrowdCanary is adaptable to changes in Twitter because it was designed based on the characteristics of users sharing information about phishing attacks, rather than using Twitter-dependent features.

\subsection{Ethical Consideration}
We took into account the ethical considerations of collecting data from Twitter on a large scale.
Although the collection and analysis targets contain massive amounts of information about Twitter accounts, the content of their tweets is public.
We believe there is no ethical issue because we did not take any actions that directly harmed users (e.g., actions on victims' email addresses or Twitter accounts).

We used common open source tools to collect data from Twitter at scale and send requests accordingly.
We conducted the experiments according to the best practices of related research on Twitter's usage guidelines, minimizing the influence on the platform.
In this experiment, we sent only 40 requests to Twitter (20 Security Keywords + 20 Co-occurrence keywords) per hour in English and Japanese.
Therefore, we believe that the availability of the platform was unaffected.

\section{Related Work}
\label{sec:relatedwork}
We describe the related research on identifying malicious tweets and generating threat intelligence from Twitter.

\noindent\textbf{Identification of Malicious Tweets.}
Numerous studies~\cite{lee2013warningbird,anupama2012phishari,gupta2018collective,thomas2011suspended,hirokicompsac} have analyzed phishing attacks that direct users to external malicious sites from Twitter.
Gao et al. proposed a system that can detect malicious posts in real time using features common to Twitter and Facebook, such as user connections and the number of characters in a post~\cite{gao2012towards}.
\textit{These studies analyze only malicious tweets (i.e., those distributed by attackers with malicious intent).
However, our study extracts benign tweets (i.e., shared by users with good intentions), and the information in the benign tweets, such as URLs or domain names, is phishing information; thus, the analysis targets are completely different.}

\noindent\textbf{Threat Intelligence Extraction from Twitter.}
Research on threat intelligence generation using Twitter information has been conducted from various perspectives~\cite{alves2020follow,khandpur2017crowdsourcing,carl2015vulnerability,hyejin2021twiti,hyejin2020cybersecurity}.
Shin et al. proposed a system to extract four types of information from a text on Twitter and external blogs: URLs, domain names, IP addresses, and hash values related to cyberattacks~\cite{hyejin2021twiti}.
It has been demonstrated that the proposed system can detect threats, especially malware-related threats, earlier than other threat intelligence systems.
Roy et al. focused on defanging and phishing attack-related hashtag strings, extracted information about phishing attacks from Twitter, and analyzed the characteristics of the accounts posting information~\cite{roy2021evaluating}.
It has been shown that information that interacts with other accounts, such as replies and retweets to the information posted on Twitter, is reflected more quickly in the block list.
\textit{Unlike our studies, tweets from security experts were collected using account names or few keywords, leading to limited Twitter data analysis.}

\section{Conclusion}
\label{sec:conclusion}
This paper proposed CrowdCanary, a system that harvests phishing information from tweets of users who have discovered or encountered phishing attacks.
The results suggest that reports from infrequent contributors (i.e., non-experts) contain a lot of valuable information for countering phishing attacks that is not included in the information posted by security experts.
Since this research showed the usefulness of information about new observation points on Twitter, we are ready to operate CrowdCanary in the future and to provide the data obtained to the national CSIRTs.
We hope that the findings of this paper will be useful for future researches and countermeasure developments.
We plan to share anonymized sample datasets with interested researchers upon request at \url{https://crowdcanary.github.io/}.

\bibliographystyle{ACM-Reference-Format}
\bibliography{main}

\end{document}